\documentclass[sigconf]{acmart}

\usepackage{tikz}
\usepackage[ruled,vlined]{algorithm2e}
\def\BibTeX{{\rm B\kern-.05em{\sc i\kern-.025em b}\kern-.08emT\kern-.1667em\lower.7ex\hbox{E}\kern-.125emX}}

\begin{document}

%
\title{How robust is MovieLens? A dataset analysis for recommender systems}

\author{Anne-Marie Tousch}
\email{am.tousch@criteo.com}
\affiliation{%
  \institution{Criteo AI Lab}
  \streetaddress{32 rue Blanche}
  \city{Paris}
  \state{France}
  \postcode{75009}
}
\begin{abstract}
Research publication requires public datasets. In recommender systems, some datasets are largely used to compare algorithms against a --supposedly-- common benchmark. Problem: for various reasons, these datasets are heavily preprocessed, making the comparison of results across papers difficult. This paper makes explicit the variety of preprocessing and evaluation protocols to test the robustness of a dataset (or lack of flexibility). While robustness is good to compare results across papers, for flexible datasets we propose a method to select a preprocessing protocol and share results more transparently.
\end{abstract}

\maketitle

\section{Introduction}

The difficulty of evaluating recommender systems is often pointed out.
Recent work showed that many papers overestimate the performance of new algorithms~\cite{Rendle2019Baselines, dacrema2019we}. 

Preprocessing also is a problem for evaluation, but the diversity of preprocessings often reflects the diversity of recommender systems applications: most datasets are private and have very different properties. 
As an example, for session-based recommender systems, researchers often use preprocessing to transform ratings datasets such as MovieLens~\cite{Harper2016ML} to this specific case.
We argue that the lack of guidelines at this step makes evaluation and comparison of algorithms harder. 

In this paper, we explicit the diversity of preprocessing protocols and use it to extract information about datasets. We analyze how metrics vary across setups.    
Our key contributions are the following:
\begin{itemize}
    \item we define a robustness metric evaluating how much a performance metric varies against preprocessing protocols of a dataset,
    \item we define a signature for preprocessed datasets,
    \item we propose a principled way of selecting a preprocessing protocol for publishing results.
\end{itemize}

\section{Method}

A recommender systems dataset preprocessing protocol involves the choice of:
\begin{enumerate}
    \item interactions preprocessing: rating thresholds, minimum number of interactions for users and items, etc.
    \item sequence processing: are timestamps used, how many events are taken as input/output, is there a maximum sequence length, etc? 
    \item training/validation/test splits and uncertainty evaluation.
\end{enumerate}

For clarity in the rest of the paper, we call a p-dataset $d$ the result of taking full, raw dataset $D$ (e.g. MovieLens-20M) and applying a preprocessing protocol.

For a given p-dataset $d$, it is straightforward to compute the performance $m(a, d)$ of an algorithm $a$ using a metric $m$. $a$ is selected from a pool of algorithms $\mathcal{A}=\{a_1, \dots, a_j, \dots a_{A}\}$. 
$m$ is taken from a pool of recommender system metrics $\mathcal{M}=\{m_1, \dots, m_i, \dots m_{M}\}$.

\subsection{Dataset robustness}\label{subsec:rob}

Given a metric $m$, the goal of the robustness is to measure how much algorithm ranks can change for this metric under variation of the preprocessing protocols. Let $\mathcal{P}(D)$ be the set of p-datasets computed from $D$. Let $m(d) = m(\mathcal{A}, d)$ be the vector of performances of each algorithm on $d$ for $m$. The \emph{robustness} is defined by:
\begin{equation}
    r(D, m) = P_{5\%}\left(\{\rho(m(d_i), m(d_j))\ |\ d_i,d_j\in\mathcal{P}(D), i<j \}\right),
\end{equation}
where $P_{5\%}$ denote the 5-th percentile of the values, and $\rho(x, y)$ is the Spearman correlation between vectors $x$ and $y$.

This metric can be interpreted easily since it is a Spearman correlation. It takes values between -1 and 1, the higher the more robust. A value of -1 means that the algorithms can be ranked in inverse order for a metric by changing the preprocessing protocol. A value of 0 means you can get completely different values. A dataset is fully robust if the rankings don't change which is reflected by a robustness of 1. Flexibility is the lack of robustness.

\subsection{Signatures and protocol selection}\label{subsec:sig}

We define the \emph{signature} of a p-dataset as the block-vector $X(d)\in\mathbb{R}^{M\times A}$ with elements $x_{i,j}(d) = m_i(a_j, d)$.

\paragraph{How to select the best protocol for my experiments?}
Practitioners typically apply an algorithm to a private dataset for production and use a public dataset for quick experimentation and publication. The best protocol is the one which allows having a public p-dataset that reproduces the private dataset $d$ properties.

The selection method is as simple as finding the public p-dataset $d^*$ from the available pool such that $X(d^*)$ is the nearest neighbor of $X(d)$.

\section{Experiments}
The dataset signature depends on the pool of metrics and algorithms. Robustness depends on the choice of a pool of preprocessing parameters, as well as metrics and algorithms.

We use classic retrieval metrics used in recommender systems: Precision@$k$, 
Recall@$k$, MMR@$k$, NDCG@$k$, as well as diversity metrics such as ItemCoverage@$k$ and APT@$k$ \cite{abdollahpouri2017controlling}, with $k=10,30,100$. We keep precise definitions of each of these metrics for the longer version of the paper.

We use five algorithms for the algorithm pool $\mathcal{A}$: Random, Best-Of (most popular items), ItemKNN, SVD and a feed-forward neural network that predicts the probability of each item to appear in the output, given the input items, following the MovieLens baseline in~\cite{Serra2017Recsys}.

We create p-datasets by changing the following parameters: rating threshold, minimum number of interactions per items, maximum number of interactions, maximum number of users, time interval between sessions, session input/output split strategy (i.e. number of "true" recommendations).

Preliminary experiments were run with the four versions of MovieLens: 100K, 1M, 10M, and 20M. We simulated four private p-datasets from two runs of reco-gym~\cite{rohde2018recogym}.

\subsection{Results}
\begin{figure}
    \centering
    \includegraphics[width=0.9\columnwidth]{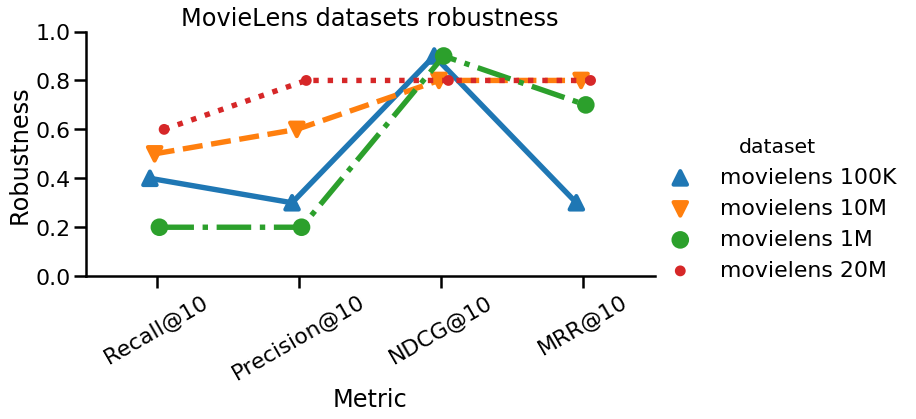}
    \caption{Robustness of MovieLens datasets against some classical metrics. The metric makes it clear the MovieLens-20M is much more robust than its smaller counterparts.}
    \label{fig:robustness}
\end{figure}

\begin{figure}
    \centering
    \includegraphics[width=0.9\columnwidth]{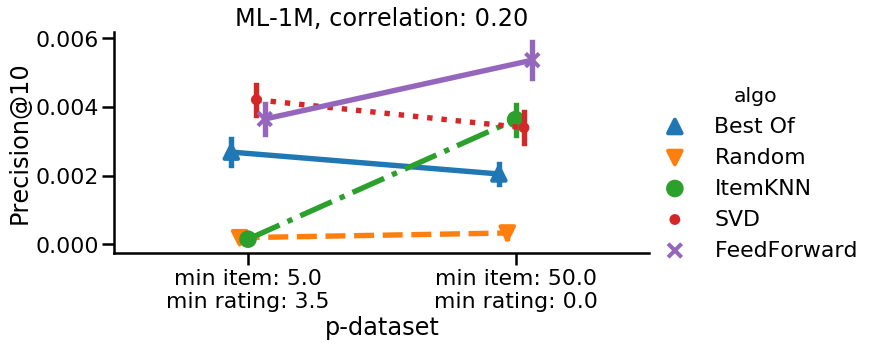}
    \caption{An example of dataset variations leading to very different metric rankings. Confidence intervals based on standard deviation on 100 bootstraps.}
    \label{fig:example}
\end{figure}

We compute the robustness of each of the MovieLens datasets. Figure~\ref{fig:robustness} shows that bigger datasets are more robust, MovieLens-20M being the most robust on average over the 4 metrics shown here. It also shows that some metrics are more robust to the changes than others: NDCG appears to be more robust than precision and recall. Figure~\ref{fig:example} gives an example of MovieLens-1M not being robust for Precision@10: it is flexible enough to make either the neural network or SVD look the best. Note that both p-datasets give very low precision.

\begin{figure}
    \centering
    \includegraphics[width=0.7\columnwidth]{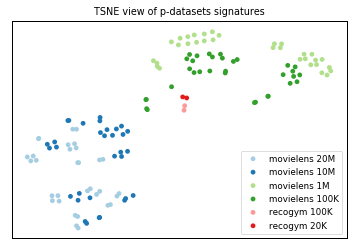}
    \caption{Embedding datasets: RecoGym datasets may be approximated by a well-chosen MovieLens counterpart.}
    \label{fig:ds_emb}
\end{figure}

Signatures can be used to visualise p-datasets with TSNE, as shown in figure~\ref{fig:ds_emb}. This could shed some light on dataset similarities. For instance, the four recogym datasets tend to be close to MovieLens-1M and 100K. In the full signature space, one of the datasets actually has a nearest neighbor computed from MovieLens-10M while the other three have matching p-datasets in MovieLens-100K. These results are encouraging, but more experiments are required to evaluate whether the performance similarity of nearest neighbors generalizes to new algorithms or metrics.

\section{Conclusion}
The importance of dataset preprocessing cannot be underestimated. If a dataset has low robustness, we show that the preprocessing can change the conclusions of the experiments. We propose a transparent method to select a protocol fitting to the target application. The longer version of this paper will include more extensive experiments to analyse the value of the proposed signature, and a study of how signature similarity generalizes to new algorithms performance similarity.

It would be ideal to have a single benchmark dataset for all recommender systems. We argue that it is unlikely that the same benchmark may cover the very different use cases of the industry and instead propose to have a transparent definition of the preprocessing protocol.

%
\bibliographystyle{ACM-Reference-Format}
\bibliography{recsys}

\end{document}